\documentclass[twoside,twocolumn,9pt]{article}
\usepackage{extsizes}
\usepackage[super,sort&compress,comma]{natbib} 
\usepackage[version=3]{mhchem}
\usepackage[left=1.5cm, right=1.5cm, top=1.785cm, bottom=2.0cm]{geometry}
\usepackage{balance}
\usepackage{mathptmx}
\usepackage{sectsty}
\usepackage{graphicx} 
\usepackage{lastpage}
\usepackage[format=plain,justification=justified,singlelinecheck=false,font={stretch=1.125,small,sf},labelfont=bf,labelsep=space]{caption}
\usepackage{float}
\usepackage{fancyhdr}
\usepackage{fnpos}
\usepackage[english]{babel}
\addto{\captionsenglish}{%
  
}
\usepackage{array}
\usepackage{droidsans}
\usepackage{charter}
\usepackage[T1]{fontenc}
\usepackage[usenames,dvipsnames]{xcolor}
\usepackage{setspace}
\usepackage[compact]{titlesec}
\usepackage{hyperref}

\usepackage{bbold,amsmath,comment,bm}

\newcommand{\tensor}{\bm}
\newcommand{\heavy}{$\mathrm{D_2O}$}
\newcommand{\normal}{$\mathrm{H_2O}$}
\newcommand{\cm}{$\mathrm{cm^{-1}}$}
\renewcommand{\vec}{\bm}
\usepackage{epstopdf}%This line makes .eps figures into .pdf - please comment out if not required.

\definecolor{cream}{RGB}{222,217,201}

\begin{document}

\pagestyle{fancy}
\thispagestyle{plain}
\fancypagestyle{plain}{
%%%HEADER%%%
\renewcommand{\headrulewidth}{0pt}
}
%%%END OF HEADER%%%

%%%PAGE SETUP - Please do not change any commands within this section%%%
\makeFNbottom
\makeatletter
\renewcommand\LARGE{\@setfontsize\LARGE{15pt}{17}}
\renewcommand\Large{\@setfontsize\Large{12pt}{14}}
\renewcommand\large{\@setfontsize\large{10pt}{12}}
\renewcommand\footnotesize{\@setfontsize\footnotesize{7pt}{10}}
\makeatother

\renewcommand{\thefootnote}{\fnsymbol{footnote}}
\renewcommand\footnoterule{\vspace*{1pt}% 
\color{cream}\hrule width 3.5in height 0.4pt \color{black}\vspace*{5pt}} 
\setcounter{secnumdepth}{5}

\makeatletter 
\renewcommand\@biblabel[1]{#1}            
\renewcommand\@makefntext[1]% 
{\noindent\makebox[0pt][r]{\@thefnmark\,}#1}
\makeatother 
\renewcommand{\figurename}{\small{Fig.}~}
\sectionfont{\sffamily\Large}
\subsectionfont{\normalsize}
\subsubsectionfont{\bf}
\setstretch{1.125} %In particular, please do not alter this line.
\setlength{\skip\footins}{0.8cm}
\setlength{\footnotesep}{0.25cm}
\setlength{\jot}{10pt}
\titlespacing*{\section}{0pt}{4pt}{4pt}
\titlespacing*{\subsection}{0pt}{15pt}{1pt}
%%%END OF PAGE SETUP%%%

%%%FOOTER%%%
\fancyfoot{}
\fancyfoot[RO]{\footnotesize{\sffamily{ ~\textbar  \hspace{2pt}\thepage}}}
\fancyfoot[LE]{\footnotesize{\sffamily{\thepage~\textbar\hspace{4.65cm} }}}
\fancyhead{}
\renewcommand{\headrulewidth}{0pt} 
\renewcommand{\footrulewidth}{0pt}
\setlength{\arrayrulewidth}{1pt}
\setlength{\columnsep}{6.5mm}
\setlength\bibsep{1pt}
%%%END OF FOOTER%%%

%%%FIGURE SETUP%%%
\makeatletter 
\newlength{\figrulesep} 
\setlength{\figrulesep}{0.5\textfloatsep} 

\newcommand{\topfigrule}{\vspace*{-1pt}% 
\noindent{\color{cream}\rule[-\figrulesep]{\columnwidth}{1.5pt}} }

\newcommand{\botfigrule}{\vspace*{-2pt}% 
\noindent{\color{cream}\rule[\figrulesep]{\columnwidth}{1.5pt}} }

\newcommand{\dblfigrule}{\vspace*{-1pt}% 
\noindent{\color{cream}\rule[-\figrulesep]{\textwidth}{1.5pt}} }

\makeatother
%%%END OF FIGURE SETUP%%%

%%%TITLE, AUTHORS AND ABSTRACT%%%
\twocolumn[
 
\noindent\LARGE{\textbf{Raman Spectrum and Polarizability of Liquid Water from Deep Neural Networks}} \\

 \noindent\large{Grace M. Sommers,\textit{$^{a*}$} Marcos F. Calegari Andrade,\textit{$^{b*}$} Linfeng Zhang\textit{$^{c}$}, Han Wang\textit{$^{d}$}, and Roberto Car\textit{$^{a,b,c\dag}$}} \\

 \noindent\normalsize{We introduce a scheme based on machine learning and deep neural networks to model the environmental dependence of the electronic polarizability in insulating materials. Application to liquid water shows that training the network with a relatively small number of molecular configurations is sufficient to predict the polarizability of arbitrary liquid configurations in close agreement with \textit{ab initio} density functional theory calculations. In combination with a neural network representation of the interatomic potential energy surface, the scheme allows us to calculate the Raman spectra along 2-nanosecond classical trajectories at different temperatures for \normal~and \heavy. The vast gains in efficiency provided by the machine learning approach enable longer trajectories and larger system sizes relative to \textit{ab initio} methods, reducing the statistical error and improving the resolution of the low-frequency Raman spectra. Decomposing the spectra into intramolecular and intermolecular contributions elucidates the mechanisms behind the temperature dependence of the low-frequency and stretch modes.} \\ \\

  ]
%%%END OF TITLE, AUTHORS AND ABSTRACT%%%

%%%FONT SETUP - please do not change any commands within this section
\renewcommand*\rmdefault{bch}\normalfont\upshape
\rmfamily
\section*{}
\vspace{-1cm}

%%%FOOTNOTES%%%

\footnotetext{\textit{$^{a}$~Department of Physics, Princeton University, Princeton, NJ 08544, USA.}}
\footnotetext{\textit{$^{b}$~Department of Chemistry, Princeton University, Princeton, NJ 08544, USA.}}
\footnotetext{\textit{$^{c}$~Program in Applied and Computational Mathematics, Princeton University, Princeton, NJ 08544, USA.}}
\footnotetext{\textit{$^{d}$~Laboratory of Computational Physics, Institute of Applied Physics and Computational Mathematics, Huayuan Road 6, Beijing 100088, P. R. China.}}

\footnotetext{\ddag~Email: rcar@princeton.edu }
\footnotetext{$^*$~Equal contribution from both authors }

%%%END OF FOOTNOTES%%%

%%%MAIN TEXT%%%%

\section{Introduction}

Raman scattering has been widely used to study rotational and vibrational spectra of gases and condensed-phase systems~\cite{Hendra1969}. The Raman effect arises from the inelastic scattering of visible light with matter, in which incident radiation is shifted in the frequency domain due to vibrations and rotations of the scatterer~\cite{mcquarrie}. Raman spectroscopy thus probes the same frequency region as infrared (IR) spectroscopy, but different selection rules apply for each technique, making Raman and IR complementary tools to investigate the rotational and vibrational signatures of condensed-phase systems.

The Raman spectrum of a system can be modeled from the polarizability time-correlation function obtained from molecular simulations. The sensitivity of the polarizability to the environmental dependence of the electronic structure demands consistent quantum mechanical approaches to model the potential and the polarizability surfaces. Traditional numerical simulations employ \textit{ab initio} molecular dynamics (AIMD), in which electronic interactions are computed on th 
13
e fly using density functional theory (DFT) and the system polarization described through the modern theory of polarization~\cite{Resta1992,King-Smith1993}. This approach was used to analyze Raman spectra of heavy water, with polarizabilities evaluated at every AIMD step using Density Functional Perturbation Theory (DFPT)~\cite{Putrino2002,raman}.
However, while AIMD methods with the appropriate functional bring much-needed predictive accuracy, their computational intensity forbids their application on large systems size and time scales~\cite{AIMD}. Empirical potentials provide a less expensive alternative, but are less robust and generalizable. One concern is that the optimal parameters for the potential energy surface do not reproduce the correct polarizability surface~\cite{empirical}; thus, successful polarizable models must take care to parameterize the polarizability surface.  
One such potential, POLI2VS, closely matches the observed IR spectrum, but the low-frequency and librational modes of the Raman spectrum are inconsistent with experiment~\cite{POLI2VS}. Moreover, non-reactive potentials are unable to model systems with mutating chemical environments, such as acidic or alkaline aqueous solutions in which proton transfer occurs on a picosecond timescale~\cite{Chen2018a}. An alternative to empirical potentials is MB-pol, a many-body potential including up to 3-body terms plus induction, parameterized with high-level quantum mechanical calculations on small molecular clusters. This is quite accurate for liquid water and more promising than empirical potentials, as it models the potential and polarization/polarizability surfaces within a self-consistent framework, but, being limited to molecular systems, this method cannot treat dissociation~\cite{MB,MBMD}.

A second consideration in modeling the Raman spectrum is the treatment of nuclear quantum effects (NQEs). Recent studies have approached the quantum TCF in several ways: approximate path integral methods such as centroid molecular dynamics (CMD) and (thermostatted) ring polymer molecular dynamics (TRPMD)~\cite{MBMD, NQE}; the local monomer approximation and other mixed quantum-classical methods, which treat exactly a small subset of vibrational modes~\cite{skinner}; and the linearized semiclassical initial value representation (LSC-IVR)~\cite{lsc-ivr}. However, each of these approximations has its drawbacks~\cite{best,rossi}, and it is difficult to distinguish effects due to approximate quantum dynamics from those due to the choice for the potential energy surface (PES).
%\RC{I changed density functional to PES to be more general}
Our goal in this paper is to demonstrate the extent to which a \textit{classical}-nuclei approach using a consistent neural-network-based PES and polarizability surface successfully reproduces experimental results.

Recently, machine learning methods have been used to express \textit{ab initio} potential energy surfaces as a function of nuclear coordinates~\cite{dpmd,behler2007,Chmiela2018,bartok2010gaussian,rupp2012fast,chmiela2017machine,schutt2017schnet,han2017deep,zhang2018end}. These methods preserve the accuracy of AIMD while improving on its efficiency.
One implementation is the Deep Potential Molecular Dynamics (DPMD), a generalizable, accurate, and linearly scalable deep neural network (DNN)-based framework that can generate long trajectories in agreement with AIMD. Some of the authors have also recently used DNNs to learn the electric polarization as a sum of local atomic contributions~\cite{zhang2020dw}. This method allows the construction of \textit{ab initio}-level polarization surfaces, which were used to compute the IR spectra of liquid and crystalline water under different pressure conditions. Alternative machine learning models based on kernel instead of DNN representations have been recently reported in the literature to predict tensorial properties extracted from \textit{ab initio} calculations~\cite{Grisafibook,Wilkins2019,Raimbault2019}. These models have been used to describe the polarizability and the Raman spectra of isolated molecules and dispersion bound molecular crystals~\cite{Raimbault2019}.

The strategy we adopt here is to use models based on DNNs to represent the quantities required by Raman spectra calculations: the interatomic potential energy and force, as well as the polarizability tensor. 
Upon training with {\it ab initio} data, DNN-based simulations reproduce AIMD results at orders of magnitude lower computational cost,
enabling accurate simulations of large-scale systems at timescales prohibitively long for AIMD. We use our DNN-predicted interatomic potential and polarizabilities to compute the Raman spectrum of liquid water at different temperatures.

Due to the wide relevance of liquid water for science and technology, experimental studies over the past several decades have investigated Raman spectra of water at different thermodynamic conditions~\cite{exp-continuum, exp-multistructure, exp-old,Walrafen1986,brooker}. Although most experiments agree on the position of primary and secondary peaks, their interpretation is still not free of controversy. For instance, it is well understood that a decrease in water temperature redshifts and broaden the primary OH stretch peak, a fact attributed to larger stability of hydrogen bonds at lower temperatures. But interpretations diverge on whether and how specific features can be assigned to unique hydrogen bond configurations. Therefore, theoretical modeling can point the way forward in identifying the physical processes at play.

\section{Methods}

In this section, we first introduce the theory of Raman spectra, relating the Raman line shape to the autocorrelator of the system polarizability. We then discuss the DNN representation of the Wannier centroid polarizabilities, before providing detail on the DFT calculations used to train the network.

\subsection{Raman Line Shape from \textit{Ab initio} Molecular Dynamics}\label{sect:theory}

The differential cross section of Raman scattering can be written in terms of the Fourier transform of the time autocorrelation function of the electronic polarizability of the system according to~\cite{mcquarrie}:
\begin{equation}
\frac{d^2\sigma}{d\omega d\Omega}(\omega) = \left(\frac{2\pi}{\lambda_s}\right)^4 \frac{1}{2\pi}\int_{-\infty}^{\infty}dt e^{-i\omega t}\langle \hat{\epsilon}_s \cdot \vec{\alpha}(0) \cdot \hat{\epsilon}_i \hat{\epsilon}_s \cdot \vec{\alpha}(t)\cdot \hat{\epsilon}_i\rangle
\end{equation}
Here $\omega$ is the Raman frequency shift, $\vec{\alpha}$ is the polarizability tensor of the sample, $\hat{\epsilon}_s$ and $\hat{\epsilon}_i$ are, respectively, the polarization directions of the scattered and incident light, $\lambda_s$ is the wavelength of the scattered light. The angular brackets denote ensemble average, and the integration is over the time $t$. Factoring out the dependence on $\lambda_s$, experiments typically report reduced line shapes $R(\omega)$ containing an arbitrary constant factor\cite{laser}:
\begin{equation}
    R(\omega) = n_{BE}(\omega)I(\omega) \propto n_{BE}(\omega)\frac{1}{2\pi}\int_{-\infty}^{\infty}dt e^{-i\omega t}\langle \hat{\epsilon}_s \cdot \vec{\alpha}(0) \cdot \hat{\epsilon}_i \hat{\epsilon}_s \cdot \vec{\alpha}(t)\cdot \hat{\epsilon}_i\rangle
\end{equation}
Here the Bose Einstein (BE) factor $n_{BE}(\omega) = 1 - \exp(-\beta \hbar \omega)$ is introduced when studying the low-frequency features of the spectrum, which would otherwise be obscured by the Rayleigh line~\cite{BE}. For fluid systems, it is convenient to decompose the polarizability tensor into a spherical part $\overline{\alpha}=\frac{1}{3}\mathrm{Tr}\tensor{\alpha}$ and a traceless anisotropic tensor $\tensor{\beta}= \tensor{\alpha} - \mathbb{1}\overline{\alpha}$, yielding the isotropic and anisotropic components of the line shape:
\begin{align}\label{eq:mcquarrie}
R_{iso}(\omega) &\propto n_{BE}(\omega) \int_{-\infty}^{\infty}dt e^{-i\omega t}\langle \overline{\alpha}(0)\overline{\alpha}(t)\rangle \notag \\
R_{aniso}(\omega) &\propto n_{BE}(\omega) \int_{-\infty}^{\infty}dt e^{-i\omega t}\frac{2}{15} \mathrm{Tr} \langle \tensor{\beta}(0)\tensor{\beta}(t) \rangle
\end{align}

In the above formulae, the polarizability depends on the nuclear coordinates, which are quantum mechanical operators but are treated here classically to compute the equilibrium time correlation functions via a molecular dynamics simulation. This amounts to neglecting nuclear quantum effects (NQEs) in the dynamics of the nuclei. In liquid water, NQEs are small but not negligible. Their influence on static equilibrium properties has been quantified in experiments and simulations based on Feynman path integrals~\cite{fluctuations}, but it is difficult to predict the effect on dynamic properties using statistical simulation methods. Leaving this issue aside, we approximate the environmental dependence of the polarizability with classical mechanics.

Nevertheless, the electronic polarizability itself can only be derived from quantum mechanics. Within the Born-Oppenheimer approximation, the electronic polarizability at time $t$ measures the response of the instantaneous polarization ($\vec{\mu}$) of the sample at time t to an infinitesimally small uniform electric field while the nuclear positions are held fixed:
\begin{equation}\label{eq:polarizability}
    \tensor{\alpha}(t) = \frac{\delta \vec{\mu}(t)}{\delta \vec{E}}
\end{equation}
We adopt first-principles density functional theory (DFT) to describe the electronic ground state of the system and use the modern theory of polarization to compute $\vec{\mu}$. The derivative in Eq.~\ref{eq:polarizability} can be expressed analytically with density functional perturbation theory (DFPT)~\cite{Baroni2001}, requiring the solution of the self-consistent response equations for the electrons, or it can be calculated numerically within the electric enthalpy framework~\cite{Umari2002,Souza2002} by applying small but finite electric fields $\pm \vec{\epsilon}$ to the sample:
\begin{equation}\label{eq:approx-polar}
\tensor{\alpha} \approx\dfrac{\vec{\mu}(\vec{\epsilon})-\vec{\mu}(-\vec{\epsilon})}{2\vec{\epsilon}}
\end{equation}
In practice, the two formulations are equivalent. We adopt here the one based on Eq.~\ref{eq:approx-polar}, which does not require a specialized DFPT code but only a DFT minimization code. The macroscopic polarization $\vec{\mu}$ of a bulk periodic system is conveniently expressed, modulo a quantum, in terms of the position vectors of the nuclei ($\vec{r}_i$) and the maximally localized Wannier centers ($\vec{w}_k$), $\vec{\mu}=e\sum_i Z_i \vec{r}_i- 2e\sum_k \vec{w}_k$, where $e$ is the unit electronic charge, $Z_i$  are atomic numbers, and we have assumed a spin-saturated system~\cite{Marzari1997}. The Wannier centers are obtained from a unitary transformation that minimizes the spatial spread in the occupied orbital subspace~\cite{Sharma2003}. We use a valence-only pseudopotential approach so that the nuclear charges $eZ_i$ are the charges of the ions consisting of the nuclei and the frozen core electrons, and the Wannier centers correspond to the valence electrons. Specializing to water, which contains oxygen ($\vec{r}_{O_i}$) and hydrogen ($\vec{r}_{H_m}$) ions, the polarization vector is:
\begin{equation}
    \vec{\mu} = 6e \sum_i \vec{r}_{O_i} + e \sum_m \vec{r}_{H_m} - 2e\sum_k \vec{w}_k
\end{equation}
In water, four Wannier centers can be uniquely associated to their nearest oxygen ion. These four Wannier centers remain close to the same oxygen during dynamical evolution, even when the water molecule to which the oxygen belongs dissociates leading to formation of hydronium and hydroxyl complexes. It is convenient to define a Wannier centroid by the average position of the four Wannier centers associated to oxygen $i$:
\begin{equation}
    \vec{w}_i = \frac{1}{4}\sum_{l_i=1}^{4}\vec{w}_{l_i}
\end{equation}
Then, the electronic polarizability of the liquid water sample is just the sum of the centroid polarizabilities $\tensor{\alpha}_i$:
\begin{equation}
    \tensor{\alpha}(t) = -8e \frac{\partial}{\partial \vec{E}}[\sum_i \vec{w}_i(t)] = \sum_i\tensor{\alpha}_i(t)
\end{equation}
In absence of molecular dissociation, the centroid polarizabilities can be viewed as effective molecular polarizabilities. These are useful for interpreting the spectrum, but only their sum is a physical observable. Its time correlation function yields the Raman line shape through Eq.~\ref{eq:mcquarrie}. The polarizability $\tensor{\alpha}(t)$ is accessible on the fly in AIMD trajectories, which thus provide a way to compute the Raman spectra using the same DFT approximation for the spectral calculations and for modeling the potential interactions that generate the atomic trajectories. Computational cost restricts these calculations to relatively short trajectories ($\approx100$ ps) and small simulation boxes ($\approx100$ molecules). The ensuing statistical errors limit the accuracy of Raman spectra calculations for water, particularly at low frequency where the Raman signal is very weak.

Recent progress with machine learning (ML) techniques applied to molecular simulation greatly alleviates these difficulties, as ML interatomic potentials can reproduce accurately AIMD trajectories at a cost that is several orders of magnitude lower than AIMD and scales linearly with system size. Our group has developed a versatile deep neural network (DNN) representation of the potential energy surface of multi-atomic systems called deep potential (DP) that has been used in several applications~\cite{dpmd,zhang2018end}. Recently, the DP representation was generalized to describe the environmental dependence of the polarization $\vec{\mu}$~\cite{zhang2020dw}. In the next section we discuss how this approach can be extended to the electronic polarizability $\tensor{\alpha}$.

\subsection{Deep Neural Network for the Environmental Dependence of the Polarizability Tensor}
Let $\mathcal{X}(\vec{r}_1,\vec{r}_2,...,\vec{r}_i,...,\vec{r}_N)$ be an extensive physical property, such as the potential energy $U$, the polarization $\vec{\mu}$, or the polarizability $\tensor{\alpha}$, which depends on the atomic positions. We consider systems for which $\mathcal{X}$ can be decomposed into a sum of local components $\mathcal{X}_i$ that depend on the coordinates of all atoms inside a finite neighborhood $\mathcal{N}_i$ of the atom at $\vec{r}_i$, i.e.:
\begin{equation}
 \mathcal{X} = \sum_i\mathcal{X}_i=\sum_i \mathcal{X}_i(\{\bm{r}_{j}\in\mathcal{N}_i\}),~\mathcal{N}_i=\{j, r_{ij}< r_c\},
  \label{eqn:w-i-NN}
\end{equation}
where $r_{ij}=||\bm{r}_i-\bm{r}_j||$ is the distance between $i$ and $j$, and $r_c$ is a predefined cut-off distance. There is no restriction, besides finite range, on the functional form of the environmental dependence in Eq.~\ref{eqn:w-i-NN}. In a condensed phase, only the global property $\mathcal{X}$ is a meaningful observable while the local quantities $\mathcal{X}_i$ are effective properties that depend on the adopted decomposition. As discussed in the previous section, in water the electronic contribution to the polarization and the electronic polarizability are conveniently given by sums of Wannier centroid contributions. Since each centroid is uniquely associated to an oxygen atom, only the oxygen neighborhoods enter the sum in Eq.~\ref{eqn:w-i-NN}. By contrast, all atomic neighborhoods, oxygen and hydrogen, are included in the sum when $\mathcal{X}$ is the potential energy $U$.

The local quantity is an effective property whose precise value depends on the adopted decomposition. 
In the case of potential energy, $\mathcal{X}_i$ denotes the atomic energy $U_i$, whose summation gives rise to the total potential energy $U$ of the system in the DNN model.
The analytical negative gradients of $U$ with respect to atomic positions define the interatomic forces.
In the case of polarizability, for the liquid water system we consider here, $\mathcal{X}_i$ becomes the molecular polarizability $\tensor{\alpha}_i$, where we only consider $i$ to be oxygen.

The environmental dependence of $\mathcal{X}_i$ can be accurately represented by a DNN-parametrized function $\mathcal{X}_i^{\vec{\gamma}}$, where $\vec{\gamma}$ denotes the parameters of the DNN model. Due to the local dependence of $\mathcal{X}_i^{\bm{\gamma}}$ on the neighborhood of $i$, the DNN model is scalable by construction. $\mathcal{X}_i^{\vec{\gamma}}$ should also satisfy some additional criteria.
First, it should depend continuously on the atomic positions and, in terms of efficiency, be orders of magnitude faster than {\it ab initio} models.
Moreover, it is crucial to preserve the translational, rotational, and permutational symmetry of the quantity that is learned.
The potential energy is a scalar quantity and should be invariant upon translation and rotation of the system and identical particle permutation. The polarization and the polarizability are vectorial and tensorial quantities, respectively, and
should be translationally and permutationally invariant, but rotationally covariant.
The above requirements are achieved by means of two DNNs, an \textit{embedding} DNN and a \textit{fitting} DNN. The embedding DNN maps the positions of the atoms belonging to the $i$th neighborhoods to features that are invariant under symmetry operations, while the fitting DNN maps these features in a way that is symmetry invariant for the potential energy, or covariant for the polarization and the polarizability. The number of hidden layers and outputs is refined in the training procedure. 

In detail, we make a \textit{local} frame transformation to the primed coordinates, which are defined relative to $\bm{r}_i$:
\begin{equation}
\bm{r}_k'\equiv\bm{r}_k-\bm{r}_i, 
  \label{eqn:r-k-local}
\end{equation}
Translational symmetry is preserved by construction in the local frame. 
It is convenient to use generalized coordinates $\bm{q}_k$, which weight atoms according to their distance from site $i$, $r_{k'}\equiv(\bm{r}_{k'}\cdot\bm{r}_{k'})^{1/2}$, and provide continuous evolution as atoms enter/exit the neighborhood.
For this purpose we introduce a smooth weight function $s(r')$ equal to $1/r'_k$ at short distances and decaying smoothly to zero as $r_k'$ approaches $r_c$, the radius of the neighborhood. 
The four-component vector
$\bm{q}_k =(q_k^1, q_k^2, q_k^3,q_k^4)$ is then given by $(s(r_k'), s(r_k')x_k'/r_k', s(r_k')y_k'/r_k', s(r_k')z_k'/r_k')$, 
in terms of the Cartesian components of $\bm{r}_k'$.
We use the matrix $\bm{Q}=(Q_{ki})=(q_k^i)$, which has $N_i$ rows and 4 columns, to represent the set of generalized coordinates $\{\bm{q}_k\}$ in a neighborhood.

The embedding DNN is the matrix $\bm{E}=(E_{ik})=E_i(s(r_k'))$ with $M$ rows and $N_i$ columns, whose elements are found by training, which maps each element in the set $\{s(r_k'),  r_k'\in\mathcal{N}_i\}$ onto $M$ outputs.
Multiplication of $\bm{E}$ by $\bm{Q}$ gives the matrix $\bm{T}=\bm{E}\bm{Q}$ with $M$ rows and 4 columns, whose generic element is:  
\begin{equation}
T_{ij}=\sum_{k=1}^{N_i}E_i(s(r_k'))Q_{kj}.
  \label{eqn:Tij}
\end{equation}
In Eq.~\ref{eqn:Tij}, the permutationally invariant sum over the atoms is a smooth function of $N_i$.
The last three columns of $\bm{T}$ ($j=2,3,4$) transform covariantly under rotation because $(Q_{k2}, Q_{k3}, Q_{k4})$ transforms like $\bm{r}_k'$. 
Let $\bm{S}$ be the matrix formed by the first $M'$ ($<M$) rows of $\bm{T}$. 
Multiplication of $\bm{T}$ by $\bm{S}^T$, the transpose of $\bm{S}$, gives the matrix $\bm{D}$ of dimension $M\times{M'}$, called the {\it{feature}} matrix:
\begin{equation}
\bm{D}=\bm{T}\bm{S}^T,
  \label{eqn:D}
\end{equation}
The elements of $\bm{D}$ are invariant under rotation and permutation. 

$\bm{D}$ captures the local features of the neighboring pattern of $i$ in a faithful and adaptive way.
While a fitting network from $\bm{D}$ to a scalar would properly define $U_i$, one needs different procedures for vectorial and tensorial quantities. The case of the polarization vector was discussed in Ref.~\citenum{zhang2020dw}. Here we specialize to the polarizability tensor. 
In this case, the output of the fitting network is an $M\times M$ $diagonal$ matrix  $\tensor{F}=(F_{jk})$,
which is mapped onto $\tensor{\alpha}_i$, in a permutationally invariant and rotationally covariant way by  right- and left-multiplications with the last three columns of $\bm{T}$ and its transpose:
\[
(\tensor{\alpha}_i)_{kl}=\sum_{j=1}^MT_{j,k+1}F_{jj}(\bm{D})T_{j,l+1}.
\]
Finally, the parameters $\bm \gamma$ are determined by training, i.e., an optimization process that minimizes a loss function, 
which is here the mean square difference between the DNN prediction and the training data. The Adam stochastic gradient descent method~\cite{Kingma2015adam} is adopted for the optimization.

The DNN for the polarizability should be combined with a DNN for the PES to study the evolution of the polarizability along MD trajectories. For consistency, the two networks should be trained with electronic structure data at the same level of theory. In this paper, we use a DNN for the polarizability and a DP representation of the PES based on the same DFT data generated with the SCAN functional approximation. Since the \textit{ab initio} electronic structure data are expensive, efficient learning strategies are crucial. To collect a minimal set of {\it ab initio} data for training, we use the iterative learning scheme Deep Potential Generator (DP-GEN) of Ref.~\citenum{zhang2019}. In this approach, an ensemble of DNN models, initially trained with a limited set of {\it ab initio} data but with different initializations of the network parameters, are used to efficiently explore the configuration space. 
A small subset of the visited configurations is selected with an error indicator, defined as the variance of the predictions within the ensemble DNNs. The protocol is repeated until all the explored configurations are described with satisfactory accuracy. The error indicator, here chosen to be the maximum standard deviation of atomic forces, exploits the highly non-linear dependence of the DNN models on the network parameters. As a consequence, different initializations of the parameters lead to different local minima in the landscape of the loss function, originating an ensemble of minimizing DNNs. In our experience, good DNN models constructed with the above procedure require significantly less \textit{ab initio} data in the target thermodynamic range than learning approaches based on independent AIMD sampling data. 

\subsection{DFT Data and Deep Neural Networks for the Potential and Polarizability of Water}

The ground state electronic structures of the equilibrium configurations of liquid water within DFT were calculated using the SCAN functional approximation~\cite{Sun2015}. SCAN predicts with sufficient accuracy the molecular structure of liquid water at equilibrium~\cite{Chen2017} as well as the dipole moment (SCAN:$1.84$~D, experiment~\cite{Clough1973}: $1.86$~D) and isotropic polarizability (SCAN: $1.41$ \AA{}$^3$ ,experiment~\cite{Murphy1977}: $1.47 $\AA{}$^3$) of the water molecule in gas phase. We used the CP code of the \texttt{Quantum ESPRESSO} package~\cite{Giannozzi2009,Giannozzi2017} to compute the electronic ground state, the potential energy, and the forces on the atoms at selected molecular configurations. The same code was also used to compute the Wannier center coordinates and the electronic polarizability at fixed nuclear positions via the electric enthalpy method~\cite{Umari2002,Nunes1994}. Norm-conserving pseudo-potentials of Troullier-Martins type~\cite{Troullier1991} were used for both oxygen and hydrogen atoms, and the wavefunctions and charge density were plane-wave expanded with an energy cutoff of $110$ and $440$~Ry, respectively. Total energy was converged to $10^{-8}$~Hartree or lower. The polarizability was estimated numerically with Eq.~\ref{eq:approx-polar} using $\epsilon=0.001$ a.u., which falls well within the linear response regime. 
An average error of $0.0005$~\AA${}^3$ for the DFT polarizabilities was estimated from a higher-order finite difference method based on a 5-point stencil. All Raman response calculations were performed at the experimental equilibrium density of water. 

\begin{figure}[h]
    \centering
    \includegraphics{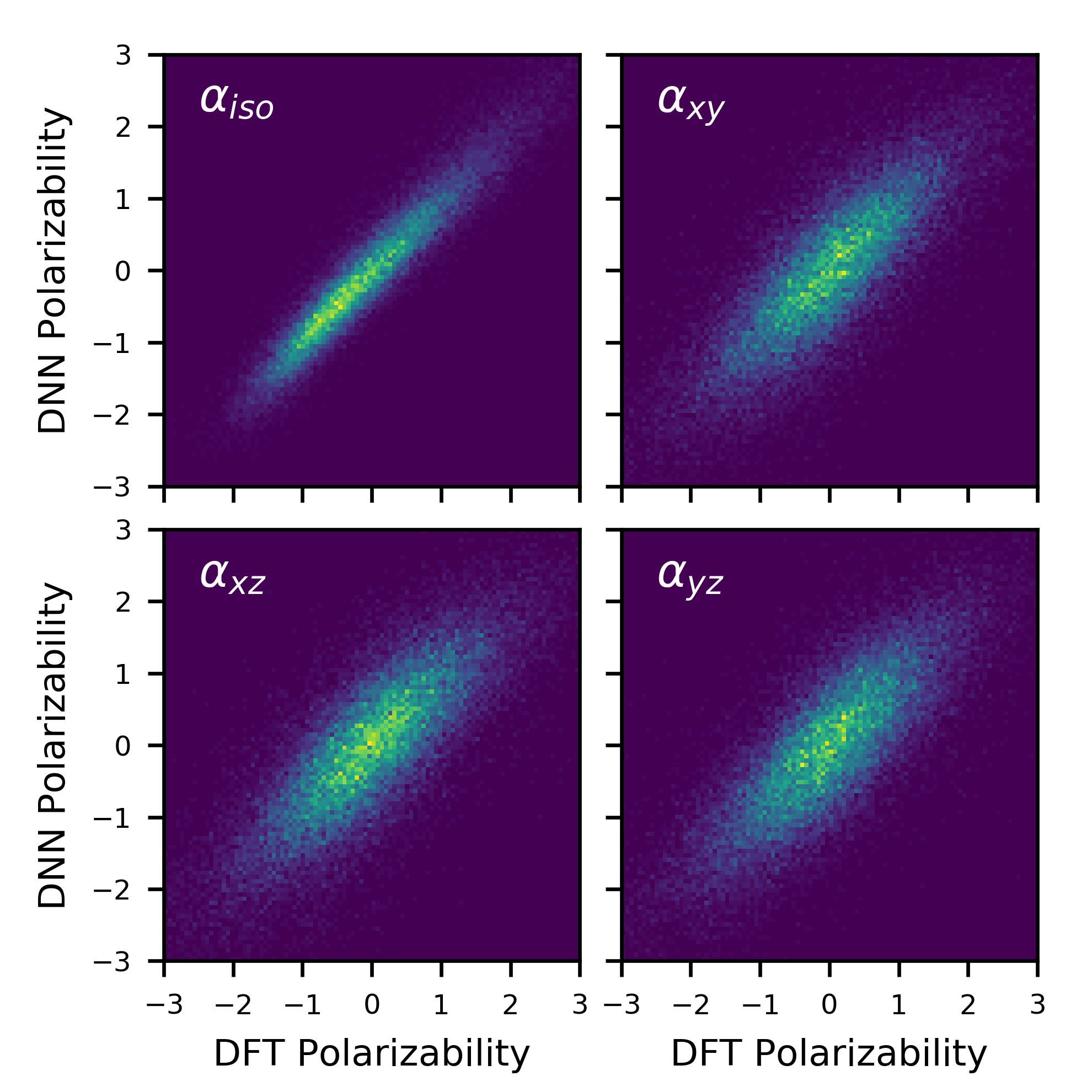}
    \caption{Water molecule effective polarizabilities in normalized units. Normalization consisted of subtraction of average DFT polarizability and division by the standard deviation of the DFT data. Isotropic polarizability and the off-diagonal components of the polarizability tensor are shown. Data compared here was not included in the training data of the deep neural network.}
    \label{fig:polar_dist}
\end{figure}

With the DP-GEN scheme we explored a pressure range of $1$--$10^5$ bar with liquid water ($192$~atoms/cell), ice Ic ($192$~atoms/cell) and ice Ih ($288$~atoms/cell). The temperature range of the exploration ranged from $270$--$370$~K for liquid water and $50$--$270$~K for ices Ih and Ic.
At the end of the iterative training procedure, our training data contained a set of $2056$, $2015$ and $1708$ configurations of liquid water, ice Ih and ice Ic, respectively. The referred training data set included only atomic forces and energy for each atomic configuration, the data needed to train the DNN potential energy surface (DP). The polarizability DNN, on the other hand, was trained only with effective molecular polarizabilities evaluated for the entire DP training set of liquid water. 
The resulting DP predicts water density ($\rho$) and diffusion coefficient ($D$) of $1.07 \pm 0.02$~g/cm$^3$ and $0.17 \pm 0.01$~\AA$^2$/ps, respectively, in close agreement with SCAN-AIMD ($\rho = 1.05 \pm 0.03$ g/cm$^3$ and $D = 0.19 \pm 0.02$~\AA$^2$/ps)~\cite{Chen2017}. 

 DP-based Molecular dynamics (DPMD) simulations of liquid water were carried out with $512$ water molecules in a periodically repeated cubic cell of $24.9$~\AA~size (H$_2$O density of $1.0$~g/cm$^3$). The system was initially equilibrated at constant volume coupled to a single Nos\'e-Hoover thermostat~\cite{Nose1984,Hoover1985} for 200~ps. 
 The simulations proceeded at constant volume and energy for 2~ns, the only section of the simulation used to compute the spectra. The classical equations of motion were integrated with the velocity-Verlet algorithm with a time step of $0.25$ and $0.5$ fs for H$_2$O and D$_2$O, respectively. All simulations were performed with the \texttt{Lammps}~\cite{Plimpton1995} package interfaced with the \texttt{DeepMD-Kit}~\cite{Wang2018}. The \texttt{DP-GEN} package~\cite{zhang2020dpgen} was used to realize the iterative learning scheme.

\subsection{Numerical Modeling of Raman Spectra}
Given the polarizability tensor as a function of time, we numerically evaluate the classical TCF of the cell polarizability and take its discrete Fourier transform (Eq.~\ref{eq:mcquarrie}). Furthermore, we also obtain the contribution of intermolecular coupling to the Raman spectra by decomposing the TCF of the cell polarizability into intramolecular and intermolecular terms.

The decomposition of the system polarizability into effective molecular polarizabilities  enables us to distinguish spectral features due to autocorrelations (within the same effective molecule) from those due to intermolecular coupling, which weaken with increasing temperature. Analogously to Wan \textit{et al.}~\cite{raman}, in evaluating Eq.~\ref{eq:mcquarrie}, we first calculated the intramolecular and intermolecular TCF. For the isotropic TCF, these are defined as:
\begin{equation}\label{eq:intra}
C_{intra}(t) = \sum_{i=1}^{N}\langle \overline{\alpha}_i(0)\overline{\alpha}_i(t) \rangle_c
\end{equation}
where $N=512$, and
\begin{equation}\label{eq:inter}
C_{inter}(t) = \sum_{i\neq j}\langle \overline{\alpha}_i(0)\overline{\alpha}_j(t) \rangle_c
\end{equation}
with analogous definitions for the anisotropic TCF. 

The decomposition of the cell polarizability into effective molecular polarizabilities also enables us to isolate sources of noise in the intermolecular spectrum. The sum in Eq.~\ref{eq:inter} was taken only over pairs $\{i, j\}$ within the first two shells of neighboring molecules (6 \AA), as determined by plotting the radial distribution function $g_{OO}(r)$. Including interactions from the third shell and beyond only adds to the zero-frequency component of the Fourier spectrum. The cutoffs were enforced on the coordinates of the oxygen atoms at each initial time used in the computation of the TCF.

\begin{figure*}
    \centering
    \includegraphics{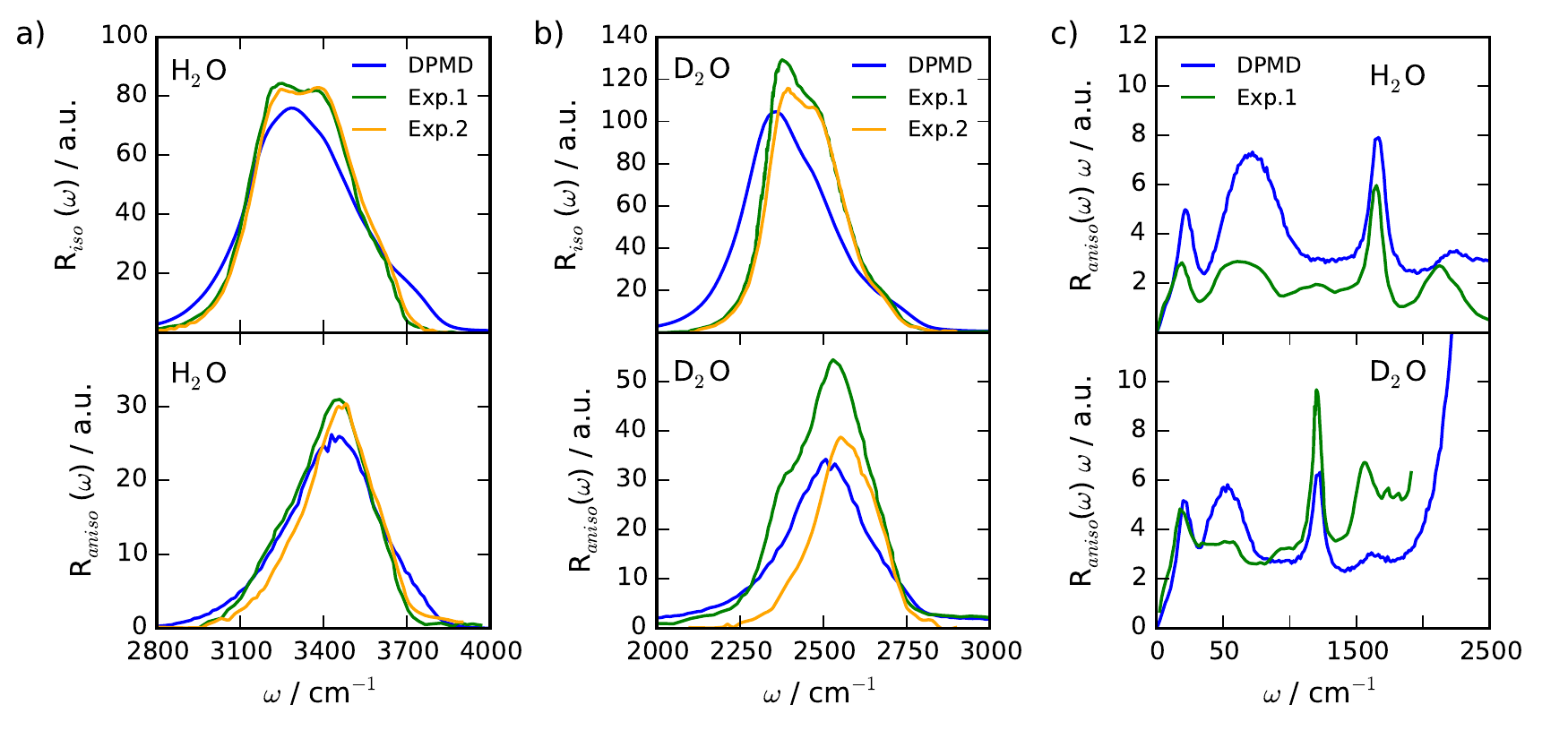}
    \caption{a) Isotropic (top) and anisotropic (bottom) Raman frequency-reduced spectra of H$_2$O at the OH stretching frequency region. b) Isotropic (top) and anisotropic (bottom) Raman spectra of D$_2$O at the OD stretching frequency region. c) Low-frequency anisotropic Raman spectra of H$_2$O (top) and D$_2$O (bottom). Intensities in (c) were divided by $1000$ for clarity. Experimental data obtained from Brooker \textit{et al.}~\cite{Brooker1989} (Exp.1, green lines) and Scherer \textit{et al.}~\cite{Scherer2005} (Exp.2, orange lines). All intensities reported in arbitrary units (a.u.). }
    \label{fig:dp_vs_exp}
\end{figure*}

Next, the spectrum was obtained by taking a discrete Fourier transform of the TCF and multiplying by the appropriate prefactors:
\begin{equation}\label{eq:me}
R(\omega_k) \propto  n_{BE}(\omega_k)\sum_{t_m=0}^{T}e^{-i \omega_k t_m} C(t_m)\Delta t \end{equation}
where $T=2.5$ ps is the length of the TCF and the times $t_m$ were discretized into intervals of $\Delta t = 0.25$ fs for \normal~and $0.5$ fs for \heavy. The frequencies $\omega_k = 2\pi k/T$, where $k$ is an integer, run from 0 to the Nyquist frequency, discretized into units of $13.33$ \cm. This provides a bound on the resolution of the spectrum, so we report the locations of peaks to the nearest 10 \cm.

\section{Results and discussion}
In this section, we present the isotropic and anisotropic Raman spectra computed from DPMD, compare to experiment, and discuss the temperature dependence of the low-frequency and stretch modes. 

We first show in Fig.~\ref{fig:polar_dist} the ability of our neural network model to predict effective molecular polarizabilities from \textit{ab initio} data. The molecular polarizabilities were obtained from a set of $416$ liquid water atomic configurations not included in the training data. Our neural network model predicts similar polarizabilities distributions to DFT, with a better agreement for the isotropic polarizability than for the off-diagonal components of the polarizability tensor. 

\subsection{Comparison with Experiment}
Our key results are shown in Fig.~\ref{fig:dp_vs_exp}, which compares the DPMD spectra to experimental data obtained from Brooker \textit{et al}~\cite{Brooker1989} and Scherer \textit{et al}~\cite{Scherer2005}. In order to properly compare the intensities from simulation and experiment, we set the integral of the DPMD isotropic spectrum of water between $2700$ and $4000$~cm$^{-1}$ equal to the integral of the same region in the experimental spectrum. The intensities reported for all the other spectra in Fig.~\ref{fig:dp_vs_exp} (except the experimental low-frequency spectra) are relative to the units defined in Fig.~\ref{fig:dp_vs_exp}a. The experimental intensities of low-frequency region were estimated by comparing the amplified and unamplified intensities of the HOH (or DOD) bending peaks reported by Brooker \textit{et al.}~\cite{Brooker1989}. 
Magnification factors of $50\times$ and $40\times$ for the amplified low frequency spectra relative to their unamplified counterparts were crudely estimated for \normal{} and \heavy{}, respectively, but this crude estimate has an error of at least $30\%$, and therefore the experimental intensities shown in Fig.~\ref{fig:dp_vs_exp}c should be taken with reservation. To make a direct comparison, we plot the temperature-reduced spectra at the OH stretch region, $R(\omega)$, and the frequency-reduced spectra, $R(\omega) \omega$, in the low-frequency region as done in Brooker \textit{et al.}~\cite{Brooker1989}

For \normal, we successfully captured the peaks in the isotropic and anisotropic spectra, at $3250$~\cm{} and $3470$~\cm{} respectively. We are also able to reproduce the accurate location of the peaks in the \heavy~ spectrum, at $\approx2350$~\cm{} for the isotropic spectrum and $\approx2530$~\cm{} for the anisotropic spectrum. The quoted wavenumbers of the peaks are defined as the positions, to the nearest $10$~\cm, of the maxima in $R(\omega)$ in the interval $2800$ to $4000$~\cm{} for \normal, and $2000$ to $3000$~\cm{} for \heavy. We also observe a shoulder at $\approx3750$~\cm{} of the \normal{} spectrum, blueshifted relative to experiment, which originates form the OH stretch of transiently broken H-bonds.
The main shortcoming of our results is the absence of a shoulder at $3400$~\cm{} in the \normal{} spectrum, previously assigned to the asymmetric stretch of water. With the exception of LSC-IVR~\cite{lsc-ivr}, this shoulder was likewise absent from the theoretical Raman spectra obtained with other methods~\cite{raman,NQE}.

\begin{figure*}
    \centering
    \includegraphics{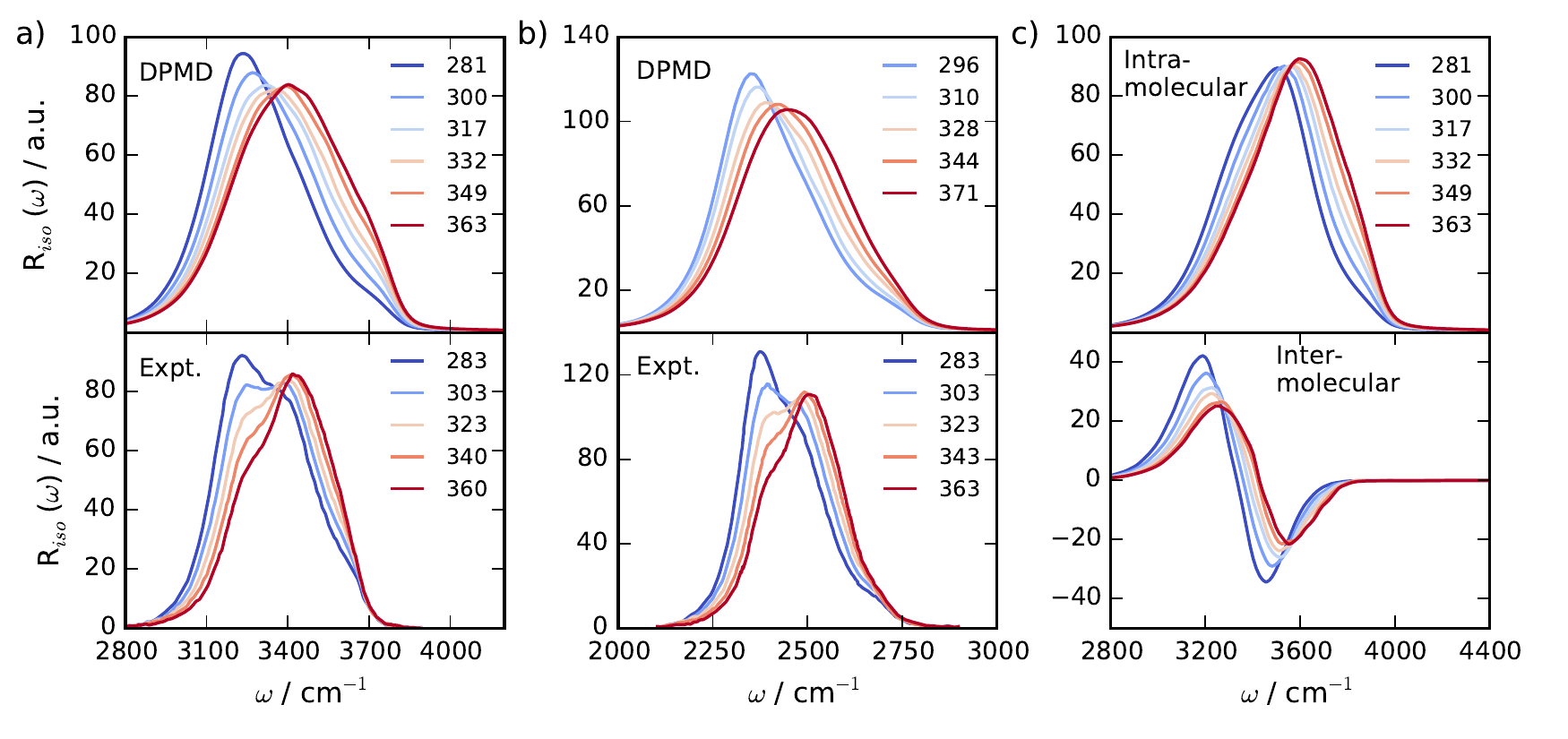}
    \caption{a) Isotropic Raman spectrum of H$_2$O at the OH stretching region as a function of temperature. b) Isotropic Raman spectrum of D$_2$O at the OD stretching region as a function of temperature. c) Intramolecular (top) and intermolecular (bottom) contributions to the isotropic Raman spectrum of H$_2$O at the OH stretching region as a function of temperature. Experimental data obtained from Scherer \textit{et al.}~\cite{Scherer2005}. Temperature reported in Kelvin units. All intensities reported in arbitrary units (a.u.).}
    \label{fig:h2o_high}
\end{figure*}

Another metric for comparison to experiment is the full width at half maximum (FWHM) of the peaks in the OH and OD stretch, which is sensitive to the choice of functional and the method of approximate quantum dynamics. Several theoretical studies have predicted an stretch band that is either too broad (such as spectra computed using the PBE functional~\cite{raman,NQE}) or too narrow (such as the \normal~spectra computed using a semiclassical approach~\cite{skinner} and MB-pol~\cite{MBMD}). We find a FWHM of $430$~\cm{} ($270$~\cm) for the isotropic OH (OD) stretch, greater than the experimental width of $418$~\cm{} ($236$~\cm) determined from the temperature-reduced spectrum in Brooker \textit{et al}. In the anisotropic spectrum, we find widths of $370$ \cm{} ($286$~\cm) for OH (OD), compared to $328$~\cm{} ($306$~\cm) from Brooker's experiment and $296$~\cm{} ($225$~\cm) from Scherer \textit{et al.}. Possible causes for this artificial broadness are the spurious electronic self-interaction present in the SCAN functional, the missing quantum fluctuations in our treatment of molecular dynamics, or a combination of both effects.

With DPMD, we are able to produce accurate results not only in the stretch band, but across the entire spectrum. We qualitatively reproduce an array of features present in the anisotropic spectra, with approximately correct intensities, over an order of magnitude less than the stretching band. These features include the network modes below $300$~\cm{}, examined in greater detail below; the libration mode from $\approx400$~\cm{} to $\approx800$~\cm{}; the bending peaks at $1670$~\cm{} and $1210$~\cm{} in \normal{} and \heavy{} respectively; and the combination bands at $1640$~\cm{} in \heavy{} and $2250$~\cm{} in \normal{}.
 The most significant discrepancy between the DPMD-predicted and experimental data is the librational peak, which extends approximately $100$~\cm{} higher in wavenumber compared to experiment. This indicates that the libration potential is stiffer than in experiment, which could result from overbonding. Moreover, our results do not capture the significant suppression of the libration peak in \heavy{} compared to \normal{} seen in the experimental data. Previous simulations, both those using the harmonic approximation~\cite{raman} and those employing approximate quantum dynamics~\cite{MBMD, NQE}, have also produced an incorrect intensity in this region. 

\subsection{Temperature Dependence of the OH/OD Stretch}

A novel contribution of this paper is that, by running the DPMD simulations at six temperatures from approximately $280$ to $370$~K, we can observe the temperature dependence of OH/OD stretch. Further, we decompose the spectra into intermolecular and intramolecular contributions, which provides insights into the mechanisms driving the temperature dependence.

The theoretical spectra in Figs.~\ref{fig:h2o_high}a and~\ref{fig:h2o_high}b exhibit a blueshift in the isotropic stretch bands due to weakening hydrogen bonds as the temperature increases. Experimentally, this blueshift manifests as a change in the relative intensities of the symmetric ($\approx3200$~\cm) and asymmetric ($\approx3400$~\cm) OH stretch peaks. The shoulder is not visible in the DPMD-computed spectra, but we do accurately reproduce the magnitude of the blueshift. Qualitatively, the experimentally observed decrease in maximum intensity is captured for both \normal ~and \heavy.

While several experimental studies have proposed Gaussian spectral decomposition of the stretch band as a mean for understanding the role of different hydrogen bond configurations, there is no consensus on the number of components~\cite{exp-multistructure}, or even whether the multistructure model is the correct starting point~\cite{exp-continuum}. \textit{Ab initio} modeling offers an alternative route to decomposing the spectrum: by splitting into the intramolecular and intermolecular contributions, we can understand how the relative intensity of couplings between and within molecules varies as the temperature increases. This is shown for the isotropic spectrum of \normal{} in~Fig.~\ref{fig:h2o_high}c.

There are several features worth noting in Fig.~\ref{fig:h2o_high}c, which together are responsible for the temperature dependence of the total spectrum. First, the intermolecular spectrum exhibits a maximum in the range of $3190$ to $3280$~\cm, followed by a minimum in the range of $3450$ to $3570$~\cm, both of approximately equal magnitude. At a given temperature, both extrema tend to redshift the frequency at which the total spectrum has a maximum. Second, as the temperature increases, both extrema are blueshifted and reduced in intensity. The same behavior occurs in the intramolecular spectrum, but to a lesser extent: the peak is blueshifted from $3430$ to $3520$~\cm{} with a slight decrease in intensity. Thus, we can attribute the overall blueshift in the spectrum not only to the separate blueshifting of the two components, but perhaps more importantly, to the reduced role of intermolecular coupling as the temperature rises. We also note that at all temperatures, the intramolecular contribution dominates.

\begin{figure}[h]
    \centering
    \includegraphics{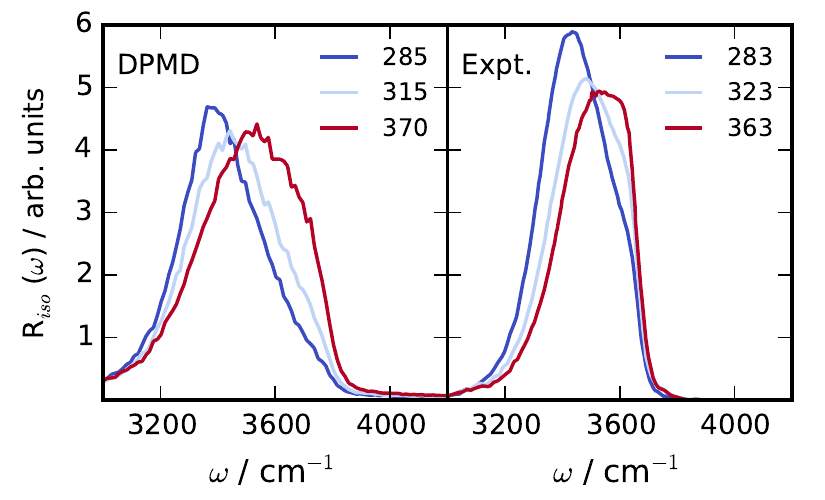}
    \caption{DPMD-predicted and experimental isotropic Raman spectra of dilute HOD in D$2$O at the OH stretching frequency. Experimental data obtained from Scherer \textit{et al.}~\cite{Scherer2005}. Intensities of the experimental spectra were normalized such that the integral of the $283$~K experimental spectrum was equal to the integral of the $285$~K DPMD spectrum between $2700$ and $4000$~cm$^{-1}$. Temperature reported in Kelvin units.}
    \label{fig:hod}
\end{figure}

The extent of intermolecular coupling in the Raman spectra of liquid water can be controlled experimentally through isotopic substitution. Raman spectra of dilute HOD in D$_2$O, for instance, is able to probe the OH stretch almost uncoupled from the surrounding oscillators. The temperature dependence of both experiment and our DPMD simulations of HOD in D$_2$O (Fig.~\ref{fig:hod}) shows a weaker temperature dependence of the OH stretch peak relative to pure H$_2$O, confirming the significance of the intermolecular coupling to the temperature dependence of H$_2$O Raman spectra. From Fig.~\ref{fig:hod} we also observe a good agreement between DPMD-predicted and the experimental spectra, although DPMD predicts broader peaks. The experimental spectra also contains a shoulder at $\approx3600$~cm$^{-1}$, usually attributed to the oscillation of non-H-bonded OH groups. This shoulder is blueshifted in the DPMD-predicted spectrum at $370$ K.

\subsection{Temperature Dependence of the Low Frequency Spectrum}

\begin{figure}[h]
    \centering
    \includegraphics{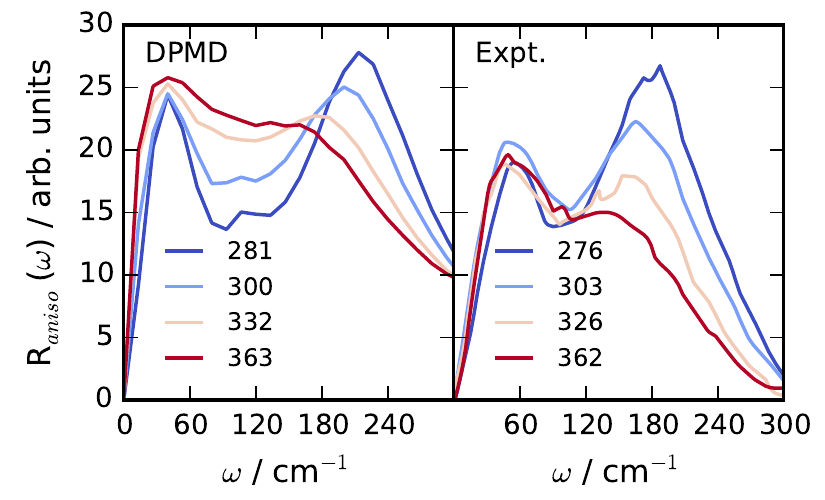}
    \caption{Low frequency region of H$_2$O anisotropic Raman spectrum as a function of temperature (K). Experimental spectra were multiplied by a constant factor in order to match maximum intensities of lowest temperature DPMD and experimental spectra. Experimental data taken from Walrafen \textit{et al.}~\cite{Walrafen1986}.}
    \label{fig:h2o_low}
\end{figure}

Turning now to the low-frequency regime, the anisotropic spectrum of \normal~below 300 \cm~is shown in Fig.~\ref{fig:h2o_low}. At 300 K, we find the nominal 60 \cm~and 180 \cm~peaks whose presence has been firmly established in experiment~\cite{Walrafen1986, walrafen2}. Walrafen \textit{et al.} further observe that as the temperature increases, the 180 \cm~peak decreases in intensity, disappearing entirely in the gas phase. This temperature dependence is evident, but slightly less pronounced in the DNN-predicted spectra relative to experiment.

While the two low-frequency peaks are experimentally attested, their origin is not entirely clear. Previous \textit{ab initio} simulations of the IR spectrum of water and ice have assigned these peaks to H bond network modes: bending of H bonded oxygen atoms ($\approx70$ \cm) and stretching of H bonded oxygen atoms ($\approx200$ \cm)~\cite{Chen2008}. Another school of thought assigns both peaks to restricted translational modes, modeling the 180 \cm~mode as a harmonic oscillator which changes from underdamped to overdamped as the temperature increases~\cite{low-freq}.

\begin{figure}[h]
    \centering
    \includegraphics{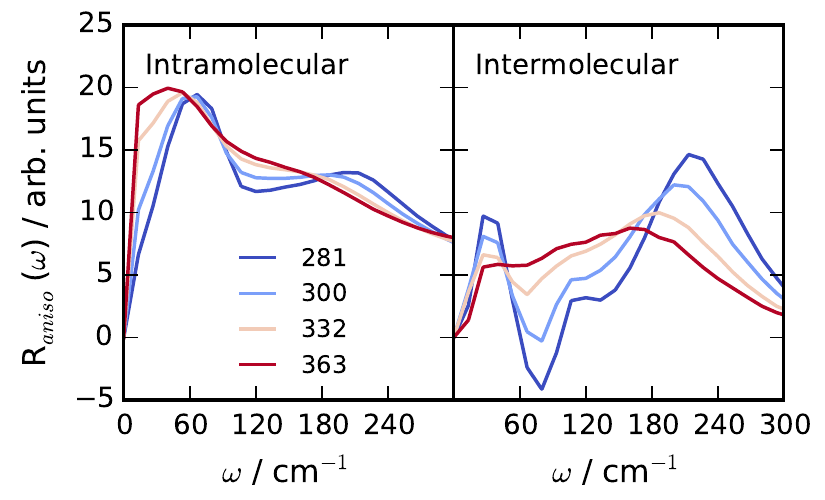}
    \caption{Intramolecular (top) and intermolecular (bottom) contributions to the low frequency region of H$_2$O anisotropic Raman spectrum as a function of temperature (K). }
    \label{fig:h2o_low_sep}
\end{figure}

In an AIMD study of the low-frequency \heavy{} spectrum, Wan \textit{et al.} attribute the $60$~\cm{} peak to intramolecular dipole induced-dipole modes and the $200$~\cm{} peak to intermolecular charge fluctuations~\cite{raman}. The DNN-predicted spectra in Fig.~\ref{fig:h2o_low_sep}, obtained from a more accurate functional, larger system size, and longer trajectories than Wan \textit{et al}, are consistent with these conclusions. At $300$~K, the nominal $60$~\cm{} peak is dominated by the maximum in the intramolecular spectrum at $70$~\cm, with a much weaker, redshifted contribution from the intermolecular spectrum at $30$~\cm{}. In addition, while the intramolecular spectrum exhibits a shoulder at $190$~\cm{}, a more distinct peak appears in the intermolecular spectrum, at $200$~\cm.

The division into intramolecular and intermolecular components also sheds some light on the temperature dependence of the peaks. As the temperature rises, the $60$~\cm{} peak in the intramolecular spectrum is slightly redshifted, and the shoulder at $210$~\cm redshifts and disappears. As in the OH stretch, the intermolecular contribution is far more sensitive to temperature; the nominal $180$~\cm{} peak redshifts from $210$~\cm{} at $281$~K to $160$~\cm{} at $363$~K, dropping in intensity by a factor of $0.6$. By accessing long time scales and controlling noise, DPMD enables us to accurately model this intermolecular contribution to the spectrum, which is crucial in explaining the temperature dependence at both low and high frequencies.

\section{Conclusion}
In this paper, we have constructed a DNN representation of the physical properties of a molecular system, focusing specifically on the effective molecular polarizability. By doing so, we were able to model the temperature dependence of the Raman spectrum, a useful tool for examining the local structure of  condensed phase systems. Liquid water offers a particularly interesting testing ground for the DNN framework, as its complex hydrogen bond networks lead to unique vibrational signatures which have thus far defied full theoretical explanation. The DPMD-predicted trends in the spectra for \normal{} and \heavy{} at temperatures ranging from $280$ to $360$ K are in good agreement with experiment, although nuclear quantum effects must be included to complete the picture.

We emphasize, however, that while the Raman spectra presented here are for water, the DNN framework is fully generalizable to other systems of interest in chemical physics. Since DPMD greatly reduces the uncertainty due to limited statistics in AIMD, it offers a useful starting point for assessing the systematic errors of DFT functionals and approximate quantum corrections relative to experiment. Although the SCAN functional describes H-bonds and intermediate van der Waals forces significantly better than GGA functionals, it is still affected by self-interaction errors. Going forward, a study of the effect of using different functional approximations in the Raman spectra will be useful. Of particular interest is the effect of functional approximations, which reduce the self-interaction errors. 
It will also be worthwhile to study how quantum corrections affect the Raman spectra. The orders-of-magnitude efficiency gains of DPMD over AIMD should enable various semi-classical methods and even permit the analytic continuation of imaginary time data.

\section*{Conflicts of interest}
There are no conflicts to declare.

\section*{Acknowledgements}
This work was conducted within the Computational Chemical Center: Chemistry in Solution and at Interfaces funded by the DoE under Award DE-SC0019394. We used resources of the National Energy Research Scientific Computing Center (DoE Contract No. DE-AC02-05cH11231). We also acknowledge use of the TIGRESS High Performance Computer Center at Princeton University. GS was supported by the PACM Summer Fellowship through the Program in Applied and Computational Mathematics at Princeton University.

%%%END OF MAIN TEXT%%%

\balance

%%%REFERENCES%%%
\bibliography{main} 
\bibliographystyle{rsc} %the RSC's .bst file

\end{document}